# Disorder-Induced Weyl Semimetal Phase and Sequential Band Inversions in PbSe-SnSe Alloys

Zhi Wang[1], Qihang Liu[2] and Alex Zunger[1,*]

[1]*Renewable and Sustainable Energy Institute, University of Colorado, Boulder, Colorado 80309, USA*

[2]*Shenzhen Institute for Quantum Science and Technology and Department of Physics, Southern University of Science and Technology, Shenzhen 518055, China*

*alex.zunger@colorado.edu

## Abstract

The search for topological systems has recently broadened to include random substitutional alloys, which lack the specific crystalline symmetries that protect topological phases, raising the question whether topological properties can be preserved, or are modified by disorder. To address this question, we avoid methods that assumed at the outset high (averaged) symmetry, using instead a fully-atomistic, topological description of alloy. Application to $(PbSe)_{1-x}(SnSe)_x$ alloy reveals that topology survives in an interesting fashion: (a) spatial randomness removes the valley degeneracy (splitting ≥150 meV), leading to a *sequential* inversion of the split valley components over a range of compositions; (b) absence of inversion lifts spin degenerates, leading to a Weyl semimetal phase without the need of external magnetic field, an unexpected result, given that the alloy constituent compounds are inversion-symmetric. (a) and (b) underpin the topological physics at low symmetry and complete the missing understanding of possible topological phases within the normal-topological insulator transition.

## Introduction

Topological insulators (TI), topological crystalline insulators (TCI) and topological Dirac semimetals (TDSM) are classified as such based on their specific band structure symmetry, *e.g.*, time-reversal symmetry[1,2] (for TI), mirror[3] or nonsymmorphic[4,5] symmetry (for TCI) and rotational symmetry[6] (for TDSM). This intimate dependence on crystalline band structure symmetry raises the question of the meaning of such topological phases in systems that are disordered alloys, for example random substitutional alloys. It has been recently suggested that in some materials, topological properties might exist in alloys. Examples include $Bi_xSb_{1-x}$[7–9] (a TI), or $(PbSe)_{1-x}(SnSe)_x$[10,11] (a TCI), or $(Na_3Bi)_x(Na_3Sb)_{1-x}$[12] (a TDSM) or $Mo_xW_{1-x}Te_2$[13] (a Weyl semimetal, WSM). A central question here is how symmetry changes with different levels of disorder, and how this affects the topological properties and the classes of topology. These questions are often overlooked, because it is sweepingly accepted in standard models of disorder such as the Virtual Crystal Approximation[14] (VCA), that a substitutional perfect random alloy should maintain the high symmetries of their underlying, non-alloyed constituent components[10,15] *i.e.* that in an $A_xB_{1-x}$ alloy, the *A* and *B* sites see an identical potential. The Single-site Coherent Potential Approximation[12,16,17] (S-CPA) makes a better approximation, distinguishing the two alloyed sublattices *A* and *B*, but assuming a that all *A* sites ( and separately all *B* sites) experience a single effective potential irrespective of the nature of its coordination by different combinations of *A* and *B* sites. In this paper we address these questions by applying a disorder model that retains full atomic resolution within density functional formalism by solving the band structure problems within large supercells, in which different local environments, as well as atomic displacements and inter-site charge transfer are allowed. The 'Effective Band Structure' (EBS) is then recovered by unfolding the supercell eigen solutions.

Application to the cubic-phase $(PbSe)_{1-x}(SnSe)_x$ TCI alloys, characterized by band edge states of the constituents that are 8-fold degenerated, reveals that:

(a) Instead of finding a *concurrent* transition between normal insulator (NI) and TCI phases, as predicted by standard alloy models[10,12,15,17] that artificially retain the high symmetry of components, we find a new, *sequential,* one-by-one inversion of the split L-valley components (by as much as 150 meV) over a range of alloy compositions. A sequential transition occurs between the insulating alloy to a metallic phase in $(PbSe)_{1-x}(SnSe)_x$ *over a finite range of composition* (here, 12% < x < 30%). This is consistent with optical experiment[18] that observe a composition range (13% < x < 24%) for the transition (however, gap smaller than 50 meV could not be detected). Perhaps future, high-resolution experiments *e.g.* low-temperature THz range optics and Angle-Resolved Photoemission Spectroscopy (ARPES) could narrow the range.

(b) The removal of spin degeneracy by lifting inversion symmetry in the alloy leads to a *Weyl semimetal phase* (the separation of Weyl points in momentum space can be larger than 0.05 Å$^{-1}$) in the sequential band inversion regime, even without the need of external magnetic field. WSM phase has recently attracted significant attention[13,19–22] because exotic physics has emerged from its surface electronic structures (*e.g.*, Fermi arc) and responses to external magnetic fields (*e.g.*, chiral anomaly). Now, a WSM has band crossings with linear dispersion (Weyl points) between two nondegenerate bands, and thus requires breaking of time-reversal or inversion symmetry. Recently, a WSM phase has been predicted to exist by tuning an order parameter (such as strain) between NI and TI phases[8,23]. However, such previous predictions suggest that the origin of Weyl phase is the broken inversion symmetry that exists already in the alloy constituents, such as $MoTe_2$ and $WTe_2$ in the $(MoTe_2)_x(WTe_2)_{1-x}$[13]. The fact that both PbSe and SnSe in cubic phase have inversion-symmetry, yet a Weyl phase is predicted in random alloy $(PbSe)_{1-x}(SnSe)_x$ is unprecedented, and implies that the effect owes its existence to inversion symmetry breaking induced by alloy disorder, not pre-existing in the constituent compounds. A recent observation[24] of linear magnetoresistance in $(PbSe)_{1-x}(SnSe)_x$ alloy in very small magnetic field reveals the symmetry breaking in charge density, which agrees with the above conclusion.

Both realizations (a) and (b) originate from the local symmetry breaking. They underpin the topological physics at low symmetry and may help us to understand the possible topological phases hidden in the transitions between NI and TI. While it has been customary to identify Weyl semimetals in structural types with broken time-reversal symmetry or broken inversion symmetry[25], the realization of Weyl phase from non-magnetic, inversion-symmetric building blocks may clarify important but missing piece in the puzzle of new topological materials.

## Results

### Symmetry and topology in substitutional random alloys.

Substitutional $(PbSe)_{1-x}(SnSe)_x$ alloys have been found long ago to show band inversion[18]. The alloy is cubic at low Sn composition, but then has a cubic-orthorhombic transition at Sn > 45%; the topological transition, whereby the cation-like conduction band minimum (CBM) and the anion-like valence band maximum (VBM) swap their order is observed around Sn = 20% *i.e.* before the cubic-orthorhombic transition[26]. What makes this system particularly interesting in the current context is that the band edges (CBM and VBM) involved in band-inversion (four L points) and a twofold Kramers-degeneracy (at each L).

Figure 1 illustrates different views on the topological transition in this alloy system. Here the cation-like, $L_6^-$ symmetric CBM (labeled $C_1$-$C_4$) is shown as blue and the anion-like, $L_6^+$ symmetric VBM ($A_1$-$A_4$) in red. The system is a normal insulator (NI) when the cation-like states are above the anion-like states (blue-above-red); otherwise, the system is a band-inverted TCI (red-above-blue). There are two possible scenarios for the topological transition: If the valley degeneracy of the L point is preserved in the alloy (Figure 1a-b), as assumed in *e.g.* VCA and S-CPA, the NI-TCI transition will involve all degenerate members getting band inverted in tandem, as a concurrent transition, and the system becomes zero gap metal at just a single composition (Figure 1b)[10,12,15,17]. Figure 1e and 1f illustrate, respectively, the VCA and S-CPA concurrent band inversions. If, on the other hand, alloy randomness splits the degenerate states, the transition occurs sequentially as shown in Figure 1c,d and extends over a range of compositions (Figure 1d). Figure 1g,h illustrates the sequential transition as found in the atomistic supercell description.

In the *monomorphous* approach it is assumed that a single structural motif tiles the entire alloy, so the alloy symmetry in perfectly random, large samples should equal that of the constituent, non-alloyed compounds. While this may be true for the *macroscopically averaged alloy configuration* $S_0 = \langle S_i \rangle$, it need not be true for any particular realization of randomly occupying the $N$ lattice sites by $A$ and $B$ atoms. Such realizations create a *polymorphous* network consisting of individual configurations $\{S_i\}$, where any of the $A_0$ sites can be surrounded locally by a different number and orientations of $A_0$-$A$ and $A_0$-$B$ bonds, and such local configurations can have not only different symmetries, but also manifest different degrees of charge transfer with respect to the neighbors, as well as different $A$-$A$, $B$-$B$, and $A$-$B$ bond lengths, all of which constitute symmetry lowering. As a result, the physical properties $P(S_i)$ of such individual, low symmetry random realizations, can be very different (*e.g.*, have splitting of degenerate levels) than the property of the fictitious, high symmetry *macroscopically averaged* configuration $S_0 = \langle S_i \rangle$. The observed physical property $\langle P \rangle$ will then be the average of the properties $\langle P(S_i) \rangle$, not the property $P(\langle S \rangle)$ of the average configuration $\langle S \rangle$. This is why alloy methods assuming that the configurationally averaged symmetry is $S_0 = \langle S_i \rangle$ predict a sharp transition between NI and TCI as illustrated in Fig 1a,b for the $(PbSe)_{1-x}(SnSe)_x$ and $(PbTe)_{1-x}(SnTe)_x$ family[10,12,15,17].

One expects that in general a random substitutional alloy will be *inherently polymorphous*, *i.e.*, each of the $A$ sites 'see' a different potential (and so do each of the $B$ sites). Whether this causes a measurable breaking of valley degeneracy depends on the delocalization vs localization of the pertinent wavefunction, which in turn reflects the physical disparity (size; electronic properties) between the two alloyed elements $A$ and $B$. We find in $(PbSe)_{1-x}(SnSe)_x$ a significant ($\geq 150$ meV) breaking of degeneracy, leading to two consequences. (a) Band inversion can happen *sequentially* (one-by-one), creating a metallic phase between the NI and TCI phases over a range of alloy compositions (Figure 1d). (b) We identify a WSM phase in the metallic regimes, which originates from the disorder-induced inversion symmetry breaking in the polymorphous network.

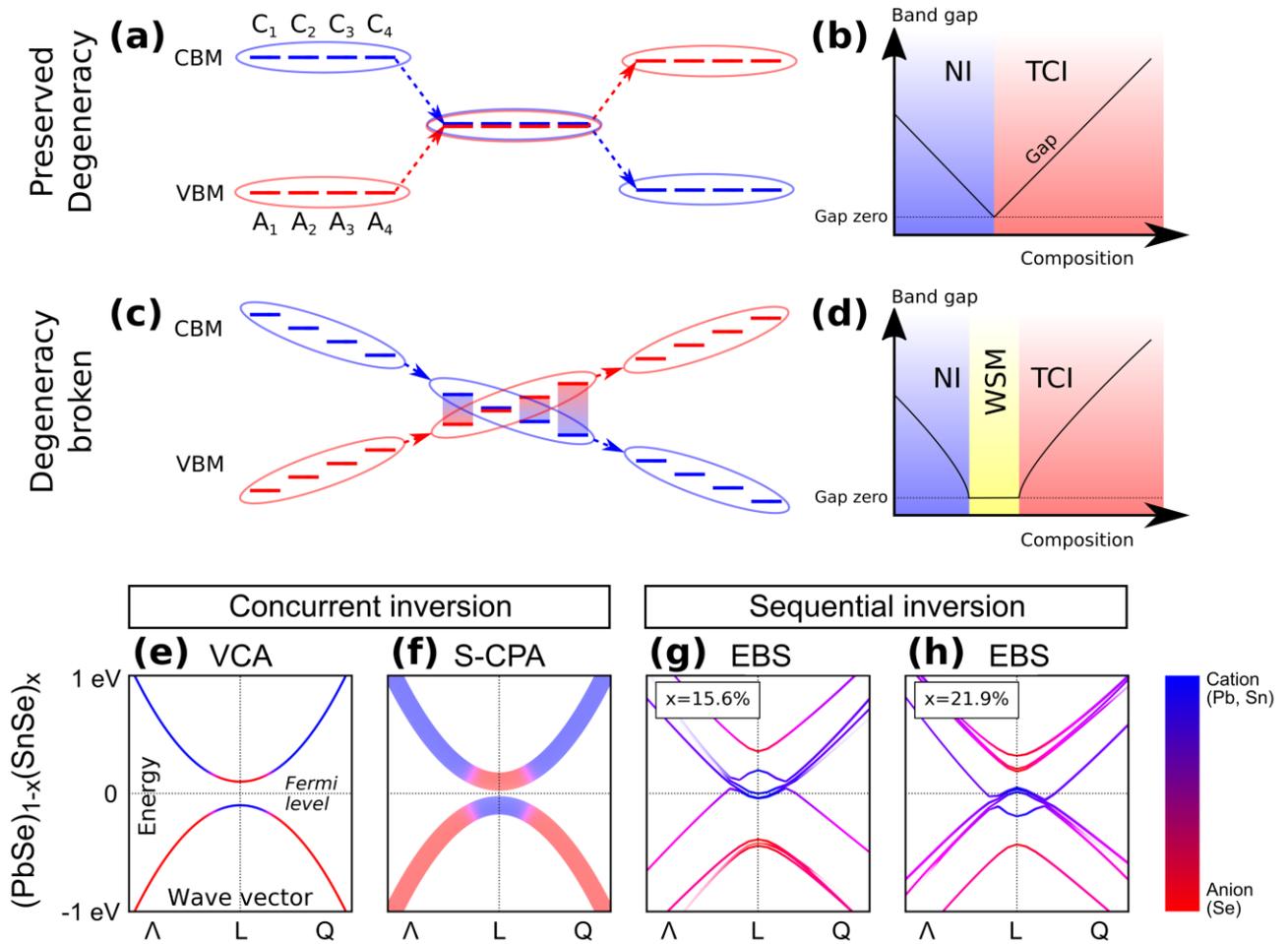

**Figure 1 | Two topological transition diagrams for (PbSe)$_{1-x}$(SnSe)$_x$ alloy.** (a) and (b): (a) shows the 'concurrent scenario' where the L valleys stay degenerate and the transition is concurrent; (b) shows the topological phase evolution with composition for the concurrent scenario. (c) and (d): (c) shows the 'sequential scenario' where randomness splits the valley degeneracy leading to sequential band inversion at L point; (d) shows the topological phase evolution corresponding to sequential scenario. The bottom row shows the alloy band structures from (e) VCA (schematic), (f) S-CPA (schematic), and (g)(h) *Effective Band Structures* (EBS) from two 256-atom *Special Quasirandom Structures* (SQS) supercells (Sn=15.6% and Sn=21.9%).

**Atomistic alloy theory with different scales of disorder.**

That the macroscopically observed property <*P*> is the average of the *properties* {$P(S_i)$}, not the property of the average configuration $P(S_0)$ is evident, among others, from the fact that local probes generally observe symmetry broken structures. Examples include the extended X-ray absorption fine structure (EXAFS) measurements of *A-X* and *B-X* bond lengths in numerous substitutional alloys such as GaAs-InAs, PbTe-GeTe, ZnTe-CdTe and PbS-PbTe[27–31], and the successful simulation of such observations on the basis of strain minimization[32]. Similarly, whereas measurements that have an intrinsically large coherence length such as X-ray diffraction tend to produce high symmetry structures when the data is fitted to small unit cell structures, more discriminating probes such as atomic pair distribution function (PDF) show alloy bond alternation and length distributions[33,34], *i.e.*, $R(A\text{-}X) \neq R(B\text{-}X)$.

To analyze the distinct physical contributions to alloy formation we consider three conceptual steps. *First,* compress the component having a larger volume and expand the one with the smaller volume, so that both can fit into the coherent alloy structure cell (here, Vegard's volume $V(x)$ for $(PbSe)_{1-x}(SnSe)_x$). Figure 2a shows that this "volume deformation" step changes linearly the gap of one component (here, SnSe) while creating a V-shape gap in another component (here, PbSe). *Second*, mix the two prepared components onto the alloy lattice at fixed volume $V(x)$, allowing charge rearrangement between the alloyed components, reflecting their possibly different Fermi levels. This is shown in Figure 2b by the self-consistent density functional theory (DFT) charge density plot in a randomly created alloy supercell. One notices a non-negligible difference in the density along the Pb-Se bond relative to the Sn-Se bond. *Finally*, given that the alloy manifests a range of different local environments (*e.g.*, the common atom Se can be locally coordinated by different metal atoms $Pb_N Sn_{6-N}$ with $0 \leq N \leq 6$), this can cause atomic displacements, illustrated here by the DFT calculated different Sn-Se and Pb-Se bond lengths at each alloy lattice constant (even if the latter follows Vegard's rule). This is illustrated in Figure 2c indicating that in the alloy environment the Pb-Se bond is distinct from the Sn-Se bond and both are different from the Vegard's average. We expect that the scale of disorder in this alloy is at an intermediate level (PbSe-SnSe mismatch of 3.5%), if compared to $(CdTe)_x(HgTe)_{1-x}$ (weakly disordered as size mismatch of CdTe-HgTe is 0.3%) and $(PbS)_x(PbTe)_{1-x}$ (strongly disordered; size mismatch of PbS-PbTe is 7.9%). The respective electronegativity difference between Pb and Sn is 0.09 (0.31 for Hg-Cd and 0.48 for S-Te).

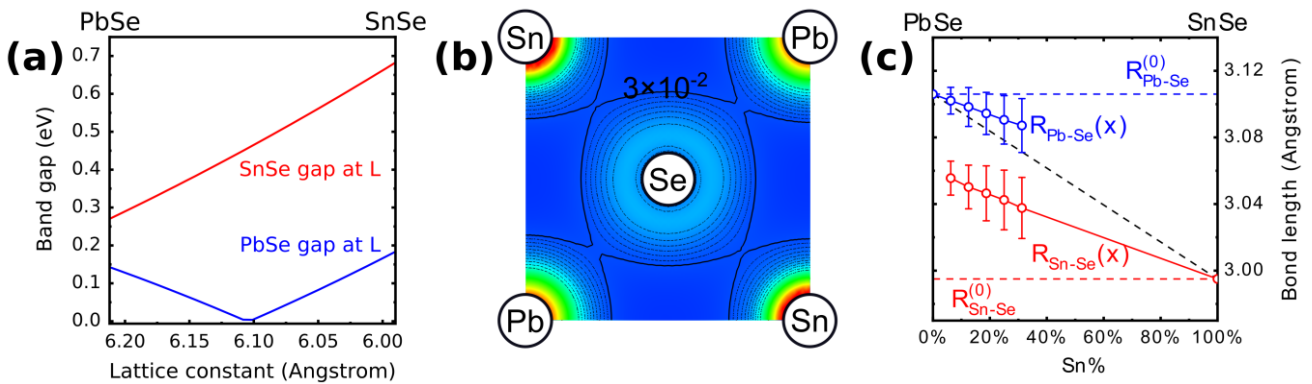

**Figure 2 | DFT results for the three leading alloy effects illustrated for $(PbSe)_{1-x}(SnSe)_x$.** (a) Band gaps at L point of PbSe (blue) and SnSe (red) under different lattice constants, (b) total charge density in $(PbSe)_{1-x}(SnSe)_x$ alloy created from equal volume components, and (c) bond length distribution of Pb-Se and Sn-Se (blue and red solid line with bars as the standard deviations) compared with their values in pure compounds (blue and red dash lines) and the Vegard's rule (black dash line), under different Sn compositions.

Monomorphous models such as VCA and S-CPA account for volume deformation but rely on the assumption that alloy charge exchange disorder or atomic displacement disorder have negligible effects (S-CPA considers approximately the former but neglects the later, while VCA neglects both) Other widely used single-site disorder methods such as the tight-binding effective Hamiltonian adds a random on-site potential $\varepsilon_i \in [-W, W]$ to the diagonal energy in the Hamiltonian[35], predicting topological Anderson insulator (TAI)[35,36], while neglecting off-site disorder (such as charge transfer and different bond lengths as shown in Figure 2b,c) and the existence of a distribution of local symmetries.

**Achieving atomic resolution for alloys in supercells**

To achieve atomic resolution of disorder one needs theory that recognizes local symmetry yet informs about the extent to which the long-range translational symmetry is retained. Such atomistic theory of alloy has been previously achieved via supercells where each $A_i$ site ($i$ = 1…$N$) has in principle a different local environment and the same for each $B_j$ site, forming a *polymorphous network* of many atomic local environments. A specially constructed *Special Quasirandom Structure* (SQS)[37,38] provides the best choice in guaranteeing the best match of correlation functions of the infinite alloy possible for $N$ atom supercell, where $N$ is increased until the required properties converge. A brief introduction of SQS has been given in Methods.

However, while supercells can directly include the atomistic effects discussed above, it folds all bands and thus results in an ensuing complex 'band structure' which is difficult to interpret as it lacks a wavevector representation (Figure 3a). Not surprisingly, the results of supercell calculations were most often presented as DOS, thus giving up the ability to recognize topological characteristics that are wavevector dependent. This absence of a relevant $E$ vs $\mathbf{k}$ dispersion relation can be solved by unfolding the bands into primitive Brillouin zone (BZ). This results in an alloy *Effective Band Structure* (EBS)[39–41] with a 3-dimentional spectral function distribution for each alloy band. Similar to angular resolved experimental spectroscopy, EBS consists of both coherent (dispersive term *e.g.* bands at L point in Figure 3b,c) and incoherent (non-dispersive broadening *e.g.* VBM at W point in Figure 3b) features, which emerges naturally from all disorder effects allowed in the supercell. With the $E$ vs $\mathbf{k}$ dispersion restored by EBS, it can be determined then if topological features are retained or destroyed. The basic concept of EBS has been given in Methods.

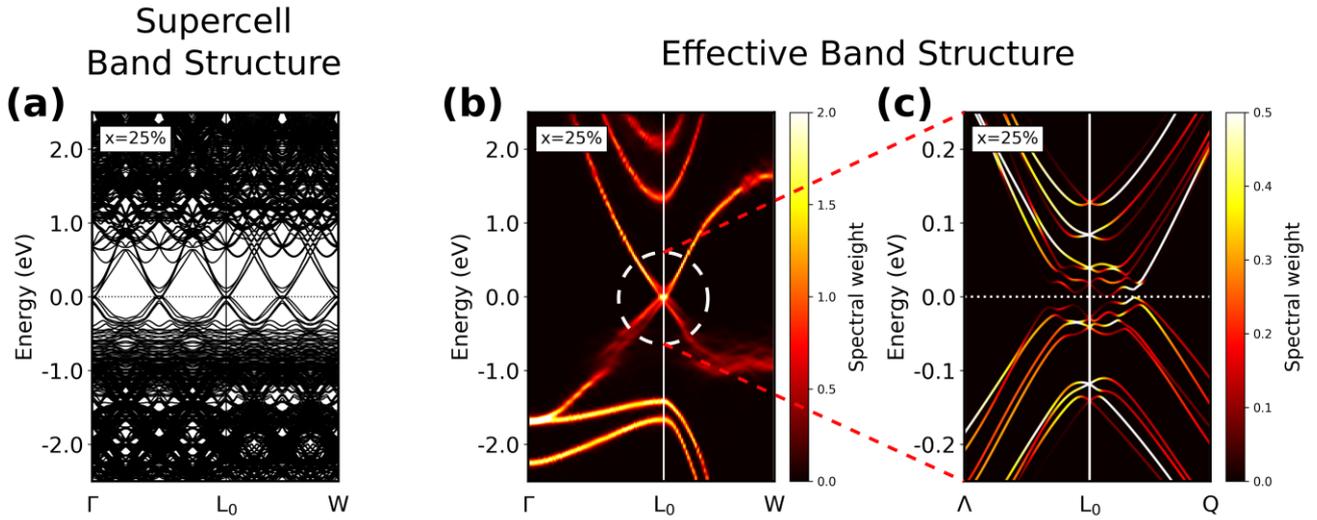

**Figure 3 | Comparison between supercell band structure and EBS in a $(PbSe)_{1-x}(SnSe)_x$ SQS supercell.** (a) Supercell band structure (256-atom supercell $(PbSe)_{1-x}(SnSe)_x$ at $x=25\%$, plotted in the primitive BZ of Fm-3m PbSe primitive cell). (b) EBS (unfolded from 256-atom supercell $(PbSe)_{1-x}(SnSe)_x$ at $x=25\%$ into the same primitive BZ as in (a)). (c) The same EBS as in (b) but zoomed-in around $L_0$ point. (a) (b) and (c) are all plotted along the same $\Gamma$-($\Lambda$)-$L_0$-(Q)-W direction in the primitive BZ ($\Lambda$ = (0.875$\pi$/a, 0.875$\pi$/a, 0.875$\pi$/a), $L_0$ = ($\pi$/a, $\pi$/a, $\pi$/a), Q = ($\pi$/a, 1.0625$\pi$/a, 0.9375$\pi$/a) and W = ($\pi$/a, 1.5$\pi$/a, 0.5$\pi$/a). As shown in (c), there are two types of splitting: valley degenerates' splitting at L point (at the vertical white solid line) which is about 150 meV, and the spin degenerates' splitting (Rashba-like) around L which is 10-30 meV.

**Broken of valley degenerates causes a sequential, one-by-one inversion of the disorder-split bands.**

We generated multiple SQS supercells for Sn composition $x$ in $(PbSe)_{1-x}(SnSe)_x$ from 0% (pure PbSe) to 31.25% to cover the NI-TCI transition regime without cubic-orthorhombic transition[26]. We found that the CBM and VBM *at L points* are not 4-fold valley-degenerated as predicted by the monomorphous approaches, instead they split and form 8 states ($C_1$-$C_4$ and $A_1$-$A_4$ as marked in Figure 1c) nearby the Fermi level at L point with a splitting energy of ≥150 meV, which reveals a significant loss of degeneracy on high-symmetry k points (here, L points). It shows that in this alloy the wavefunction indeed feels the alloy disorders, and the single band picture from monomorphous description is inadequate. The splitting of degenerates in $(PbSe)_{1-x}(SnSe)_x$ has been shown in Figure 3c.

We then made a statistical analysis among different SQS (grouped by the Sn composition) on the eigen energies of the 8 split bands at L, which is shown in Figure 4. We found that the 8 bands have relatively stable valley splitting and small overlap, making the band inversion at L point a sequential process, *i.e.*, at low composition the band inversion firstly occurs between $C_4$ and $A_1$, then with composition increasing the other bands get inverted one-by-one. This sequential inversion regime emerges from the lifting of the of valley degeneracy at L by the alloy disorder. Therefore, this regime is not observable in monomorphous approaches, where the NI-TCI transition is sharp and concurrent.

The band gap always locates between the 4th and 5th bands of the 8 split bands, however because of the one-by-one band inversion (*e.g.*, at some composition the 4 unoccupied bands are $C_1$, $A_1$, $C_2$ and $A_2$, while the 4 occupied bands are $C_3$, $A_3$, $C_4$ and $A_4$), band gap in this regime is very small hence can be absent (beyond the measurable range of equipment)[18] in measurement. In optical experiment[18], the positive-gap (normal insulating) composition range was x < 10%, and composition range where gap < 50 meV ( the detection limit) was 13% < x < 24% . Our results suggest a sequential band inversion in the composition range of 12% < x < 30%. Besides a possible DFT error, the comparison between our method and experiment could also be affected by (1) the absence of band gap closing points in experiment (instrument could not measure gap in far-infrared with IR detectors at the time of the experiment), and (2) non-randomness effect and high carrier concentration in experimental samples. Nevertheless, the existence of the composition range over which the transition occurs, and the stable splitting energy suggest that this sequential inversion regime might be observable in high-resolution experiments *e.g.* low-temperature THz range optics and Angle-Resolved Photoemission Spectroscopy (ARPES).

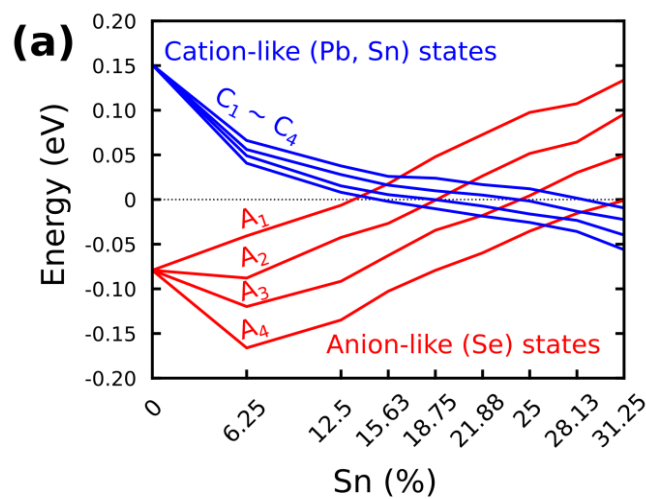

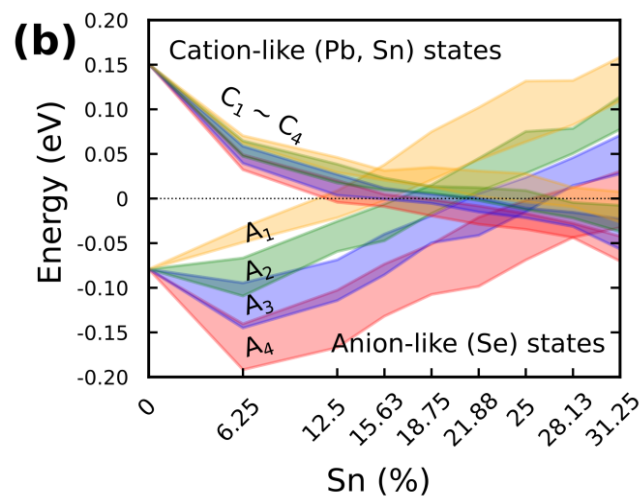

**Figure 4 | Sequential band inversion and band broadenings in $(PbSe)_{1-x}(SnSe)_x$ supercells.** (a) The averaged eigen values of the 4 cation-like (blue, $C_1$-$C_4$) and 4 anion-like (red, $A_1$-$A_4$) band branches from EBS at L point in primitive first BZ at different Sn compositions, calculated statistically from 160 SQS supercells (all are 256-atom supercells). (b) The standard deviations of the eigen values of $C_1$-$C_4$ and $A_1$-$A_4$, calculated statistically from 160 SQS supercells.

**Broken inversion symmetry due to alloy disorder leads to Weyl semimetal phase in the sequential band inversion regime.**

The complex band crossing shown in Figure 3 and 4 inspired us to study the topological property hidden inside the sequential band inversion regime. By calculating the band gap, we found that the polymorphous approach predicts metallic phase (bulk gap equals to zero) in the regime of the sequential band inversion process, where the four $C_1$-$C_4$ and the four $A_1$-$A_4$ have crosses among each other (12% < x < 30%). Within this metallic phase the mirror Chern number does not apply to characterize NI or TCI transition. We found that $(PbSe)_{1-x}(SnSe)_x$ alloys have bulk Weyl points in this sequential band inversion regime, which drives the system to a WSM. The Weyl points originate directly from the broken of local inversion symmetry, which is attributed to two reasons: (a) the atomic potentials are different between Pb and Sn, and (b) the atomic displacements are polymorphous and non-uniform. The intensity of inversion symmetry broken can be seen from the spin degenerates' splitting (10-30 meV) around L point, as shown in Figure 3c. Note that previous works indicated that Weyl phases can appear between NI and TCI or TI phase[42,43], however they used either external magnetic field or non-centrosymmetric compounds, thus the time reversal symmetry or inversion symmetry has been broken by external knob or already in the selected building blocks. The conclusions in previous works are hence not applicable to our system, where the constituent compounds are both time-reversal and inversion symmetric.

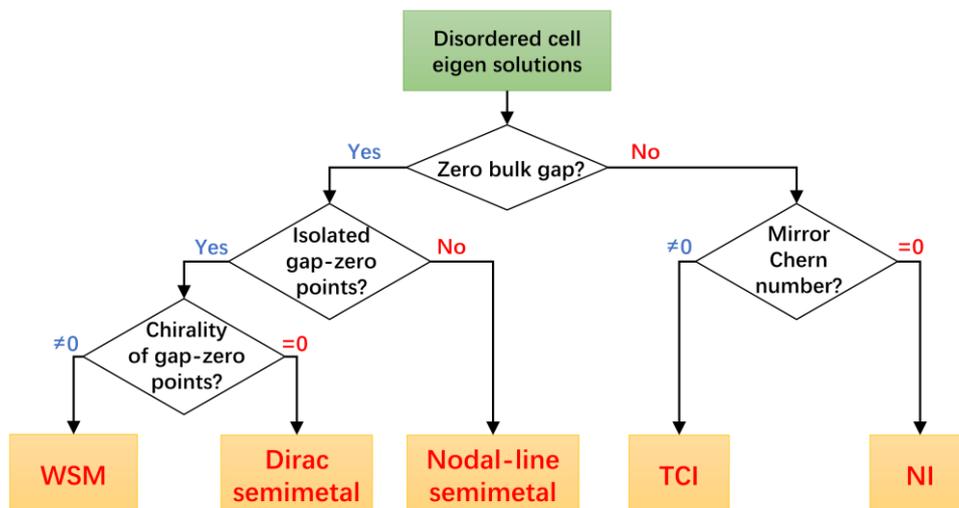

**Figure 5 | The work flow chart for the recognition of topological phases.**

By choosing different sizes and different space groups of the alloy supercell, we systematically look into the emergent phases in the band version regime. The Sn composition of each size of supercell has been fixed to 25%. We then use only uniform hydrostatic pressure, *i.e.*, changing lattice constant continuously, as an order parameter to tune the band inversion and thus the NI-TCI topological transition in each supercell. Note that the motivation to change the lattice constant is to simulate the phase diagram with the concentration of Se. Łusakowski *et al.*[44] predicted a metallic phase in $(PbTe)_{1-x}(SnTe)_x$ alloy, also using uniform hydrostatic pressure as the order parameter, where they found many 'jumps' of the topological invariant as a function of order parameter in a 16-atom highly symmetric supercell. They then predicted that this metallic phase would also exist in larger 64- and 216-atom supercells (about 1000 configurations in total). Unfortunately, they did not examine if the metallic phase is topological, or symmetry- or size-related.

Figure 5 shows our flow chart for the recognition of topological properties in our systems. The results have been listed in Table I. We started from the highly symmetric, 8-atom supercells (space group *Pm-3m*), and removed symmetries from supercell step by step, by enlarging the size and using atomic displacements (from *Pm-3m* to *Amm2*, to *Pmm2*, to *PM*, to *P1*). We found that (1) the metallic, sequential band inversion regime exists in every supercell we calculated, and (2) the zero bulk band-gap always occurs on points or lines in momentum space. Furthermore, these gap-zero points or lines have symmetry-related properties: in the 8-atom supercells (*Pm-3m*), gap-zero points are Dirac points, protected by the inversion symmetry; removing symmetries step by step drives the gap-zero points from Dirac points to nodal-lines (*Amm2*), and finally to Weyl points (*Pmm2*, *PM*, and *P1*). Notice that the *P1* symmetry supercells are no longer TCI (due to the broken of mirror plane), but they still have the WSM phase in the sequential inversion regime.

**Table I | The summary of topological phase transition in different supercells of $(PbSe)_{0.75}(SnSe)_{0.25}$.** We show in the first row the prediction from VCA and S-CPA as a comparison.

| Method | Space group | SG index | No-inversion regime | Sequential band inversion regime | Full-inversion regime |
|---|---|---|---|---|---|
| VCA, S-CPA | *Fm-3m* (2 atoms) | 225 | NI | *No such regime* | TCI |
| Supercells | *Pm-3m* (8 atoms) | 221 | NI | Dirac semimetal | TCI |
| | *Amm2* (8 atoms) | 38 | NI | Nodal-line semimetal | TCI |
| | *Pmm2* (24 atoms) | 25 | NI | WSM | TCI |
| | *Pm* (48 atoms) | 6 | NI | WSM | TCI |

| | | | | |
|---|---|---|---|---|
| *P1* (32 atoms) | 1 | NI | WSM | NI |

We further showed the Weyl points in momentum space in Figure 6: we chose the supercells having 24 atoms and space group *Pm* (lowest symmetry that can be TCI) at different lattice constants, then plotted the shortest distance of Weyl points that have opposite chirality ('Weyl pair'), as well as the trajectories of Weyl points, as variables of lattice constant. We observed 4 Weyl points in this kind of supercell, forming two Weyl pairs. The separation between Weyl pair can be larger than 0.05 Å$^{-1}$, which is equal to the predicted value in (PbSe)$_{1-x}$(SnSe)$_x$ alloy with external magnetic field[42]. Moreover, the sequential band inversion regime (marked in yellow in Figure 6a) always has non-zero separation between Weyl pairs. It illustrated that WSM phase exists not accidently but in a wide range within the sequential band inversion range.

The appearance of Weyl phase in sequential inversion regime explains the unusual oscillation in the spin Chern number noted by Łusakowski *et al.*[44] Furthermore, a recent observation[24] of linear magnetoresistance in (PbSe)$_{1-x}$(SnSe)$_x$ alloy in very small magnetic field also reveals the symmetry breaking in charge density. We show that Weyl phase can appear in alloys even with its components having both time-reversal and inversion symmetries, which cannot be predicted if one assumes that the random alloy always restores the symmetry of constituent compounds. Such Weyl phase originates from the local symmetry breaking induced by disorder, which agrees with the magnetoresistance experiment and reveals the important role of polymorphous network in the topological transition. For a random alloy, it is expected that different local configurations with disorder-induced symmetry breaking manifest Weyl points forming a spot rather than a point in the momentum space. We also expect that other alloy systems, *e.g.*, halide perovskites *ABX*$_3$, with bigger size mismatch between atoms will have larger atomic displacements, cause larger removal of inversion symmetry, and hence have Weyl points easier to be measured from experiments.

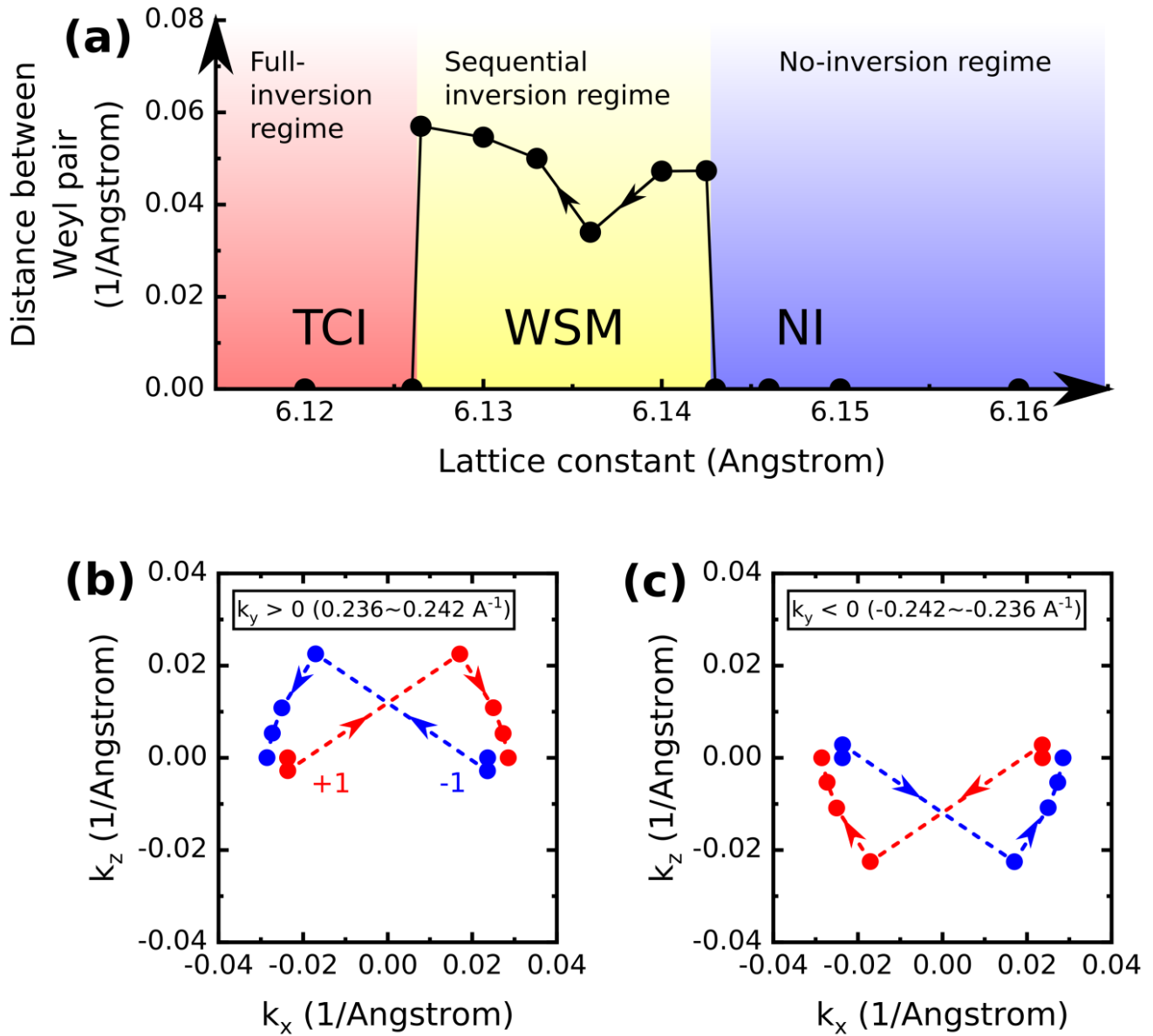

**Figure 6 | The separations and trajectories of Weyl points in k-space in (PbSe)$_{0.75}$(SnSe)$_{0.25}$ supercells.** All supercells are 24-atom with space group *Pm*. (a) The shortest distance in k-space between Weyl points having opposite chirality, as a variable of lattice constant. (b) (c) The trajectories of the 4 Weyl points (which have been grouped into 2 pairs: one in k$_y$ > 0 region while another one in k$_y$ < 0 region) in k-space, at different lattice constants. The directions of k$_x$, k$_y$ and k$_z$ are (1-10), (110) and (001) directions in cubic phase, respectively. Mirror plane is k$_x$=0. The origin points (0, 0, 0) in (b) and (c) are L point. Red and blue circles in (b) and (c) are Weyl points with chirality +1 and -1, respectively. The arrows in (b) and (c) are corresponding to the arrows in (a).

## Discussion.

In the (PbSe)$_{1-x}$(SnSe)$_x$ alloy we predicted a stable splitting (≥ 150 meV) of valley degenerates at L point, and a metallic, sequential band inversion regime between NI and TCI phases in a range of compositions, both of which are absent in conventional monomorphous alloy theories. Furthermore, we predicted this sequential band inversion regime to be WSM phase because of symmetry lowering by alloy disorder, where the separation between Weyl points in momentum space can be larger than 0.05 Å$^{-1}$. External magnetic field can then enhance, but is not essential for, this WSM phase. We suggest that the sequential band inversion can be measured in high-resolution experiments, and that the Weyl semimetal alloy from time-reversal-symmetric and inversion-symmetric building blocks may expand our horizon for the design and search for new topological phases.

# *Methods.*

### Computational setups.

VASP package[45] has been used in all calculations. We use the Wannier90 code[46] and WannierTools code[47] to solve the surface states of supercells. The EBS has been done using a modified version of the BandUP code[48]. A Dudarev type[49] U = 2 eV has been applied on Pb s-orbital[50] in all calculations. The lattice constant of Fm-3m PbSe compound is taken from DFT calculation, while the lattice constant of Fm-3m SnSe is taken from experiment. Table MI shows the comparison between DFT and experimental results[18]. We use a 2 × 2 × 2 k-mesh for 256-atom supercells, and an 8 × 8 × 8 k-mesh for primitive cells to perform the BZ integrations. Energy cutoff of 360 eV, total energy convergence of $10^{-7}$ eV/f.u. and force tolerance of $5 \times 10^{-3}$ Angstrom/eV (if bond relaxation is allowed) have been chosen in all cases. For the supercell relaxation, we fix the cell size and shape following Vegard's rule, then relax all internal atomic positions.

**Table MI | Comparison between DFT and experiment results for pure compounds PbSe and SnSe.**

|  | DFT | | | Experiment | |
| --- | --- | --- | --- | --- | --- |
|  | **Method** | Gap (eV) | Lattice constant (A) | Gap (eV) | Lattice constant (A) |
| PbSe ($Fm\bar{3}m$) | PBE-GGA+U*, SOC | 0.23 | 6.212 | 0.17 (77 K), 0.23 (195 K), 0.27 (300 K) | 6.12 |
| SnSe ($Fm\bar{3}m$) | PBE-GGA, SOC | 0.72 | 5.99 | 0.62-0.72** | 6.00 |

*U = 2eV on Pb s-orbital
**Estimated from straight line fittings[18]

### SQS: introduction, generation and convergence tests.

SQS is designed to find a single realization in a given supercell size to best reproduce the properties in infinite alloy. As the pair, 3-body, 4-body *etc.* correlation functions can all be calculated precisely in the perfectly random, infinite alloy, SQS then searches all possible configurations in the *N*-atom supercell to find the best correlation functions compared to the ones in infinite alloy. Therefore, a property *P* calculated from an SQS is not simply a single 'snapshot' but approximates the ensemble average <*P*> from many random configurations. Description and discussion of SQS can be found in Ref[37,38]. Ref[38], furthermore, showed that large size SQS gives more reliable results than calculating ensemble averages directly from many small random supercells, because some intermediate range interactions (*e.g.* long-range pairs) in large supercells do not exist in small ones due to size limitation. Note that convergence of *P* as a function of SQS size must be tested before one applies such SQS in calculations.

The mixing enthalpy of (PbSe)$_{1-x}$(SnSe)$_x$ alloy system is very low so we consider only perfect randomness and no phase separation in alloy. Although there is a transition from cubic to orthorhombic at Sn=45%, because the experimental NI-TCI transition composition is around Sn=20%[26], we only consider the cubic phase. Meanwhile, in the cubic regime the alloy lattice constant shows a good linearity with Sn composition[51].

We constructed different sizes of supercells for the (PbSe)$_{1-x}$(SnSe)$_x$ alloy. For the study of valley degeneracy splitting and sequential band inversion, we used 256-atom SQS supercells. According to our convergence tests, the 256-atom supercells while considering pair and triplet correlation functions have stable band gaps. We do not introduce any artificial off-center displacements because the 256-atom supercells naturally have polymorphic, (n)-dependent atomic relaxation. In Figure M1 we show the convergence tests of band gaps using different SQS supercell sizes and different cutoff distances for 2- and 3-body correlation functions. We suggest that a 256-atom supercell SQS with a cutoff distance of 2.13-time lattice constant is good enough to mimic the perfectly random alloy. For each Sn composition from Sn = 6.25% to 31.3% (8 compositions in total), we have generated 20 256-atom NaCl-structure SQS supercells, *i.e.*, there are 160 256-atom SQS supercells in total. The reason we use multiple supercells for single composition is to investigate the consistency of band inversion among different SQS realizations.

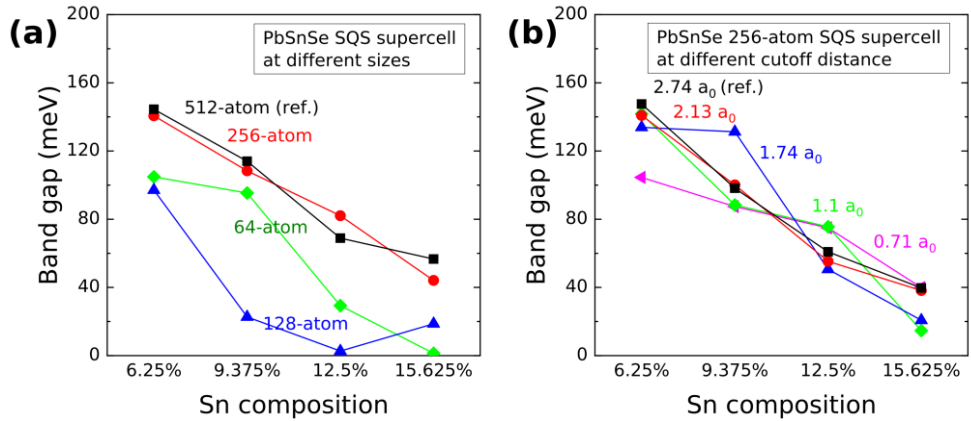

**Figure M1 | Convergence test for (PbSe)$_{1-x}$(SnSe)$_x$ SQS supercells.** Band gap at L point (a) when increasing the size of SQS supercell, and (b) when increasing the distance cutoff of 2- and 3-body correlation functions in 256-atom SQS supercell. In (a) the gaps from 512-atom SQS supercells are chosen as references ('ref.'), while in (b) the gaps from 2.74-$a_0$ (lattice constant) cutoff distance are chosen as references.

For the study of WSM phase and Weyl points, we constructed different supercells with target symmetries by (1) first generating multiple SQS supercells with given size then (2) picking up the cells that have target symmetries. For each row in Table I we calculated one supercell, changing its lattice constant continuously, as an order parameter to tune the band inversion. The searching of Weyl points and the calculation of chirality of Weyl points are then done in the WannierTools code.

**Effective Band Structure**

The basic concept of EBS can be described using the following equations. Assume in supercell $|Km\rangle$ is the m-th electronic eigen state at $K$ in supercell BZ whereas in primitive cell $|k_i n\rangle$ is the n-th eigen state at $k_i$ in primitive BZ, then each $|Km\rangle$ can be expanded on a complete set of $|k_i n\rangle$ where $K = k_i - G_i$, and $G_i$ being reciprocal lattice vectors in the supercell BZ, which is the folding mechanism[41]

$$|Km\rangle = \sum_{i=1}^{N_K} \sum_n F(k_i, n; K, m)|k_i n\rangle, \quad (1)$$

The supercell band structure at $K$ can then be *unfolded* back to $k_i$ by calculating the spectral weight $P_{Km}(k_i)$

$$P_{Km}(k_i) = \sum_n |\langle Km|k_i n\rangle|^2 \quad (2)$$

$P_{Km}(k_i)$ represents 'how much' Bloch characteristics of wavevector $k_i$ has been preserved in $|Km\rangle$ when $E_n = E_m$. The EBS is then calculated by spectral function $A(k_i, E)$

$$A(k_i, E) = \sum_m P_{Km}(k_i)\delta(E_m - E) \quad (3)$$

# References.

**Acknowledgment.**

The work at the University of Colorado at Boulder was supported by the National Science foundation NSF Grant NSF-DMR-CMMT No. DMR-1724791. The ab-initio calculations were done using the Extreme Science and Engineering Discovery Environment (XSEDE), which is supported by National Science Foundation grant number ACI-1548562. We thank Dr. Quansheng Wu for fruitful discussions.

**Author contributions.**

Zhi Wang performed all DFT and postprocessing calculations, as well as theoretical analysis. Alex Zunger performed analysis of the results and directed the writing of the manuscript with contributions and discussion from all. Qihang Liu contributed to the theoretical analysis and discussion on topological phase and phase transition.

**Competing interests.**

The authors declare no competing interests.